\documentclass[twocolumn,showpacs,preprintnumbers,amsmath,amssymb]{revtex4}
\usepackage{graphicx}
\usepackage{dcolumn}
\usepackage{bm}
\usepackage{color}
\usepackage[applemac]{inputenc}

\begin{document}
\title{On the interpretation of negative birefringence observed in strong-field optical pump-probe experiments: high-order Kerr and plasma grating effects}
\author{G. Karras}
\author{P. B\'ejot}
\author{J. Houzet}
\author{E. Hertz}
\author{F. Billard}
\author{B. Lavorel}
\author{O. Faucher}

\affiliation{Laboratoire Interdisciplinaire Carnot de Bourgogne (ICB), UMR 6303 CNRS-Universit\'e de Bourgogne, 9 Av. A. Savary, BP 47 870, F-21078 DIJON Cedex, France}

\newlength{\textlarg}
\newcommand{\strike}[1]{%
  \settowidth{\textlarg}{#1}
  #1\hspace{-\textlarg}\rule[0.5ex]{\textlarg}{0.5pt}}

\date{\today}
%42.65.Jx Beam trapping, self focusing and defocusing, self-phase modulation; 52.38.Hb Self-focusing, channeling, and filamentation in plasmas; 42.65.Re	Ultrafast processes;
%37.10.Vz	Mechanical effects of light on atoms, molecules, and ions
%42.65.Re  	Ultrafast processes
% 42.65 Hw : Phase conjugation; photorefractive and Kerr effects

\begin{abstract}
The analysis of negative birefringence optically induced in major air components (Loriot \textit{et al.}, \cite{Loriot17_2009,Loriot18_2010}) is revisited in light of the recently reported plasma grating-induced phase-shift effect predicted for strong field pump-probe experiments (Wahlstrand and Milchberg, \cite{Wahlstrand2011OPL}). The nonlinear birefringence induced by a short and intense laser pulse in argon is measured by femtosecond time-resolved polarimetry. The experiments are performed with degenerate colors, where the pump and probe beam share the same spectrum, or with two different colors and non-overlapping spectra. The interpretation of the experimental results is substantiated using a numerical 3D+1 model accounting for nonlinear propagation effects, cross-beam geometry of the interacting laser pulses, and detection technique. The model also includes the ionization rate of argon and high-order Kerr indices introduced by Loriot \textit{et al.} enabling to assess the contribution of both terms to the observed effect. The results show that the ionization-induced phase-shift  has  a minor contribution compared to the high-order Kerr effect formerly introduced, the latter allowing a reasonably good reproduction of the experimental data for the present conditions.
\end{abstract}

\maketitle
\section{INTRODUCTION}
\label{introduction}
Recently, we reported on the observation of unexpected large nonlinear dynamics in the measurements of optically induced birefringence of major air components using a pump-probe time-resolved polarization technique \cite{Loriot17_2009,Loriot18_2010}. For the investigated N$_2$, O$_2$, and Ar gases, the intensity dependence of the reported effect shared the same behavior. In the lower limit, i.e., below $\sim$10 TW/cm$^2$, the induced birefringence followed the applied field intensity. Around 25 TW/cm$^2$, a saturation of the signal was observed leading to an inversion and then followed by a very fast change of the birefringence for larger intensities. We attributed the transition between these linear and nonlinear regimes to a saturation of the electronic Kerr effect induced by the intense short laser pulse. Based on different assumptions, we proposed a phenomenological model by introducing in the standard definition of the nonlinear refractive index $n_{\textrm{Kerr}}=n_2 I$, where $n_2$ is the Kerr index and $I$ is the applied intensity \cite{Boyd2008}, high-order Kerr (HOK) terms $ n_4$, $n_6$, $n_8$, and eventually $ n_{10}$ \cite{Loriot17_2009,Loriot18_2010,Loriot2011}. Starting from the weak field regime, each term was successively fitted to the experiment by gradually increasing the intensity so as to minimize cross correlations. The intensity at the interaction region was carefully estimated by simultaneously recording the retarded birefringence signal resulting from  field-free  alignment of a molecular gas \cite{Stapelfeldt75_2003,Seideman52_2006,Kumarappan76_2007}. In the HOK  model, where the nonlinear refractive index is defined by $n_{\textrm{Kerr}} = n_2 I +n_4 I^2+ n_6 I^3+ n_8 I^4$, the saturation and inversion of the electronic Kerr index are due to negative $n_4$ and $n_8$ terms.
It is well known that the optical Kerr effect plays a central role in the nonlinear propagation of short and intense laser pulses such as self-phase modulation, self-focusing, pulse compression, and laser filamentation \cite{Bejot81_2010,Schmidt96_2010,Couairon_Mysy_2007,Berge_wolf_2007,Chin_report_2005}. The potential impact of the HOK  model on the last process was presented in Refs. \cite{Bejot104_2010} and \cite{Loriot2011}. These works reveal that Kerr saturation due to high-order terms can lead under certain conditions to a self-guided nonlinear pulse propagation without requiring the need for a plasma production. This prediction stands against the standard model of filamentation where the plasma acts as the primary mechanism responsible for the arrest of Kerr-induced self focusing in a gas. It is important to mention that the HOK model predicts also the standard regime of filamentation; the boundary between the standard and the all-Kerr regime being ruled by the characteristics of the applied laser electric field \cite{Loriot2011}. The existence of these two distinguishable filamentation regimes was supported by an experiment \cite{Bejot106_2011} that compared long pulse versus short pulse filamentation produced in argon using a bipulse setup.

The change of paradigm introduced by the HOK  model \cite{Polynkin2011} has ignited an ongoing active debate questioning the physical meaning or the existence of the Kerr saturation reported in Refs. \cite{Loriot17_2009,Loriot18_2010}. In this respect, several theoretical studies have been focused on the microscopic origin of the HOK effect  either by solving the 1D or 3D Schr\"odinger equation for an atom exposed to a strong laser field \cite{Kano_2006,Nurhuda66_2008,Nurhuda10_2008,Teleki82_2010,Volkova_2011,Bandrauk2012,Bejot110_2013,Bandrauk2013} or by applying the concept of nonlinear Kramers-Kronig relations \cite{Bre106_2011,Wahlstrand2012PRL}. Although the literature published on this issue has contributed to get a much better insight into the temporal dynamics responsible for the nonlinear atomic response at the driving field frequency, the relevance of the HOK model for short-pulse long-wavelength propagation is still controversial. The existence of the HOK effect has also been challenged by the recent report of Wahlstrand and Milchberg on ionization-induced birefringence in pump-probe experiments \cite{Wahlstrand2011OPL}. The authors of this work argue that a pump-probe cross-phase shift generated through a plasma grating (PG) could provide an alternative explanation to the negative birefringence observed in Refs. \cite{Loriot17_2009,Loriot18_2010}. In more details, they show that in case of non-collinear pump-probe experiment employing degenerate frequencies (i.e., the same color for the pump and probe beam) and non-collinear linearly polarized fields, the optical interference between the pump and probe field leads to a spatial intensity grating. When the pump intensity is sufficiently high, the ionization induced by this intensity modulation embeds a plasma grating into the gas medium. Because the grating is due to the addition of the pump field with the projection of the probe field along the direction of the pump field, the resulting two-beam coupling effect only affects the phase of the parallel field components. This leads to a transient birefringence undergone by the probe beam during the temporal overlapping of the two beams. The pump-probe signal predicted by the plasma grating is featured by an asymmetric profile as well as a power dependence proportional to the ionization rate of the gas \cite{Wahlstrand2011OPL,Wahlstrand85_2012}. The former lies in the fact that the plasma stems from the accumulation of ionized electrons during the pulse and therefore behaves as a retarded effect.

The aim of the present work is to assess the role of the plasma grating on the negative birefringence signal  observed by Loriot \textit{et al.} in a pump-probe experiment \cite{Loriot17_2009,Loriot18_2010}. To this end, we will present a quantitative analysis of a new series of experimental data based on 3D+1 numerical simulations including both semi-empirical HOK and PG models. By investigating the influence of the pulse parameters like the pump intensity,  frequency chirp, and probe frequency we will show that the negative birefringence observed in the experiment can not be solely ascribed to the plasma grating  effect. The experimental setup used for the one- and two-color transient birefringence measurements is discussed in section \ref{setup_section}. The theoretical model used for the numerical simulations of the 3D+1 nonlinear propagation in a pump-probe cross-beam geometry is described in section \ref{model}. The results of the one- and two-color experiment together with direct comparisons with numerical simulations are presented and discussed in section \ref{results_section}.

\section{EXPERIMENTAL SETUP}
\label{setup_section}
 \begin{figure}[tb!]
  \begin{center}
    \includegraphics[keepaspectratio, width=8cm]{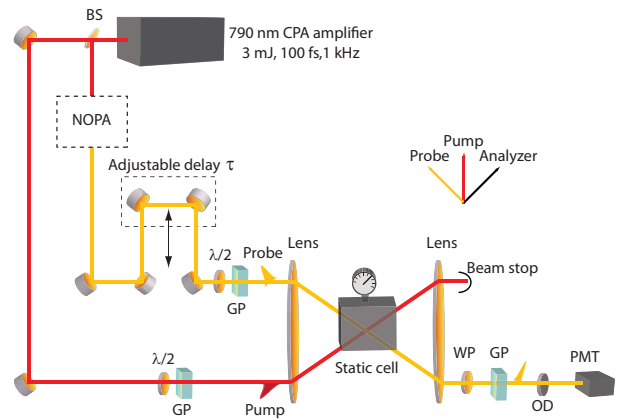}
  \end{center}
  \caption{(Color online) Pump-probe setup for laser-induced birefringence measurements. BS: Beam Splitter, GP: Glan Polarizer, OD: Neutral Optical Density, WP: Wave Plate, PMT: Photo Multiplier Tube. The relative polarizations of the pump, probe, and analyzer are shown in the inset. The noncollinear optical parametric amplifier (NOPA) is only used for the purpose of the two-color pump-probe measurements.}
  \label{setup}
\end{figure}
The experimental setup used for the presented measurements is depicted in Fig. \ref{setup}. It is a typical pump-probe scheme that has been reported previously \cite{Loriot17_2009} and takes advantage of the depolarization of the probe beam when the later interacts with a birefringent medium optically induced by a pump pulse. The laser source is a chirped pulse amplified Ti:Sapphire system which delivers pulses centered at 790 nm with 3 mJ energy per pulse (100 fs pulse duration) at 1 kHz repetition rate. Pump and probe pulses are formed by splitting the output beam into two parts using a beam splitter. Control of the energy and polarization of each pulse is accomplished using zero-order half-wave plates and glan polarizers, while their relative delay is adjusted using a motorized stage. Wavelength tunability, here for the probe beam, whenever needed is attained by directing the beam to a non-collinear optical parametric amplifier (NOPA). The intensity ratio of the probe over the pump beam is less than 1$\%$ and they are both linearly polarized at 45$^\circ$ relative to each other. Both beams are overlapped at small angle (3$^\circ$) and focused in the center of a 27 cm-long static cell  with a f=30 cm plano-convex focal lens resulting in  beam waists of 40 $\mu$m. The gas used for the reported data is Ar and the pressure in the cell is kept low in order to minimize the contribution from nonlinear propagation effects. The depolarized probe beam after the cell passes through a linear crossed analyzer and the signal is measured using a photomultiplier and sampled by a boxcar integrator. This homodyne signal carries the information about the transient induced birefringence and can be written as
\begin{eqnarray}
\mathcal{S}\propto\int_{-\infty}^{\infty} I_{\textrm{probe}}\left(t-\tau\right)\left[\Delta n\left(t,I_{\textrm{pump}}\right)\right]^2dt,
\end{eqnarray}
where $\Delta n$ is the difference in the refractive index along and perpendicular the laser polarization axis and $I_{\textrm{probe}}$ ($I_{\textrm{pump}}$) is the intensity of the probe (pump) beam. From the above equation we see that the sign of the transient birefringence is not revealed in this case due to the quadratic dependence of the signal on $\Delta n$. In order to gain access to the sign of this transient birefringence we add a local oscillator (L.O.) using an adjustable wave plate. The amplitude of this permanent birefringence is adjusted  by changing the applied mechanical stress and the sign  is  controlled by rotating the wave plate perpendicular to the propagation direction by 90$^\circ$   \cite{Lavorel2000,Loriot2007}. The detected signal in this case is given by
\begin{eqnarray}
\mathcal{S}_{\pm}\propto\int_{-\infty}^{\infty} I_{\textrm{probe}}\left(t-\tau\right)\left[\Delta n\left(t,I_{\textrm{pump}}\right)\pm\textrm{L.O.}\right]^2dt,
\label{local osci}
\end{eqnarray}
where $\pm$ denotes the sign of the L.O. Subtraction of the two experimental traces for positive and negative L.O. results to the pure heterodyned signal:
\begin{eqnarray}
 S_{\textrm{heterodyne}}\propto\left(\mathcal{S}_+-\mathcal{S}_-\right).
\end{eqnarray}
As discussed in section \ref{results_section}, this procedure allows to prevent the detection  of possible energy transfer or  dichroism.

\section{NUMERICAL MODEL}
\label{model}
The equations system driving the propagation of the global electric field envelopes $\mathcal{E}_{x}$ ($\mathcal{E}_{y}$) polarized along $x$ ($y$) read in the reciprocal space \cite{Kolesik_PRE}:

\begin{eqnarray}
\partial_z\widetilde{\mathcal{E}}_x&=&i\left(k_z-k_1\omega\right)\widetilde{\mathcal{E}}_x+\frac{1}{k_z}\left(\frac{i\omega^2}{c^2}\widetilde{P}_{x,\textrm{NL}}-\frac{\omega}{2\epsilon_0c^2}\widetilde{J}_x\right)\nonumber\\
& & -\widetilde{L}_{x,\textrm{losses}},\\
\partial_z\widetilde{\mathcal{E}}_y&=&i\left(k_z-k_1\omega\right)\widetilde{\mathcal{E}}_y+\frac{1}{k_z}\left(\frac{i\omega^2}{c^2}\widetilde{P}_{y,\textrm{NL}}-\frac{\omega}{2\epsilon_0c^2}\widetilde{J}_y\right)\nonumber\\
& & -\widetilde{L}_{y,\textrm{losses}},
\end{eqnarray}
where $\widetilde{P}_\textrm{NL}$ is the  nonlinear polarization, $\widetilde{J}$ is the plasma-induced current, $\widetilde{L}$ accounts for  nonlinear losses expressed in the reciprocal space, and
\begin{eqnarray}
k(\omega)&=&n(\omega)\omega/c,\nonumber\\
k_z&=&\sqrt{k^2(\omega)-(k_x^2+k_y^2)},\nonumber\\
P_{x,\textrm{NL}}&=&n_{2}|\mathcal{E}_x|^{2}\mathcal{E}_x+\frac{2}{3}n_{2}|\mathcal{E}_y|^{2}\mathcal{E}_x+\Delta n_{x,\textrm{HOK}}\mathcal{E}_x,\nonumber\\
P_{y,\textrm{NL}}&=&n_{2}|\mathcal{E}_y|^{2}\mathcal{E}_y+\frac{2}{3}n_{2}|\mathcal{E}_x|^{2}\mathcal{E}_y+\Delta n_{y,\textrm{HOK}}\mathcal{E}_y,\nonumber\\
\widetilde{J}_{x,y}&=&\frac{e^2}{m_\textrm{e}}\frac{\nu_\textrm{e}+i\omega}{\nu_\textrm{e}^2+\omega^2}\widetilde{\rho\mathcal{E}}_{x,y},\nonumber\\
\partial_t\rho&=&\sigma_N\left(|\mathcal{E}_x|^{2N}+|\mathcal{E}_y|^{2N}\right)(\rho_\textrm{at}-\rho),\nonumber\\L_{x,\textrm{losses}}&=&\frac{N\hbar\omega_0\sigma_N\rho_\textrm{at}}{2}|\mathcal{E}_x|^{2N-2}\mathcal{E}_x,\nonumber\\
L_{y,\textrm{losses}}&=&\frac{N\hbar\omega_0\sigma_N\rho_\textrm{at}}{2}|\mathcal{E}_y|^{2N-2}\mathcal{E}_y,\nonumber\\
\Delta n_{x,\textrm{HOK}}&=&\sum_{j=2}^{5}{n_{2j}\left(|\mathcal{E}_x|^{2j}+\frac{j+1}{2j+1}|\mathcal{E}_y|^{2j}\right)},\label{delta_n_x}\\
\Delta n_{y,\textrm{HOK}}&=&\sum_{j=2}^{5}{n_{2j}\left(|\mathcal{E}_y|^{2j}+\frac{j+1}{2j+1}|\mathcal{E}_x|^{2j}\right)},\label{delta_n_y}
\end{eqnarray}
where $\omega$ is the angular frequency, $n$ is the frequency dependent refractive index, $c$ is the light velocity, $k(\omega)$ is the frequency dependent wavenumber, $k_1=\partial k/\partial \omega|\omega_0$ corresponds to the inverse of the group velocity, $k_x$ and $k_y$ are the conjugate variables of $x$ and $y$ in the reciprocal space, $m_\textrm{e}$ and $e$ are the electron mass and charge respectively, $n_2$ is the nonlinear refractive index, $\rho$ ($\rho_{\textrm{at}}$) is the electron (atom) density, $N$ is the ionization nonlinearity, $\sigma_N$ is the ionization cross-section, and $\nu_\textrm{e}$ is the effective collision frequency (see Table \ref{phys_para}). In Eqs. \ref{delta_n_x} and \ref{delta_n_y},  $n_{2j}$ are the high-order Kerr terms measured in \cite{Loriot17_2009,Loriot18_2010}.
\begin{table}[htbp!]
 \begin{tabular}{|c|c|c|c|}
 \hline
 n$_2$ & N & $\sigma_N$ & $\nu_e$ \\
 \hline
 10$^{-7}$ cm$^2$/TW/\textrm{bar} & & 10$^{-3}$ TW$^{-N}$cm$^{2N}$/s/\textrm{bar} & fs$^{-1}/\textrm{bar}$ \\
 \hline
 1 & 7.5 & 1.26 & 190\\
 \hline
 \end{tabular}
 \caption{Physical parameters used in the model: Kerr index $n_2$ \cite{Loriot18_2010},   nonlinearity $N$ and  ionization cross-section $\sigma_N$  \cite{KasparianChin00,Loriot41_2008}, and effective collision frequency $\nu_e$  of Ar at 800 nm.}
 \label{phys_para}
\end{table}
 \begin{figure}[tb!]
 \begin{center}
    \includegraphics[keepaspectratio, width=8cm]{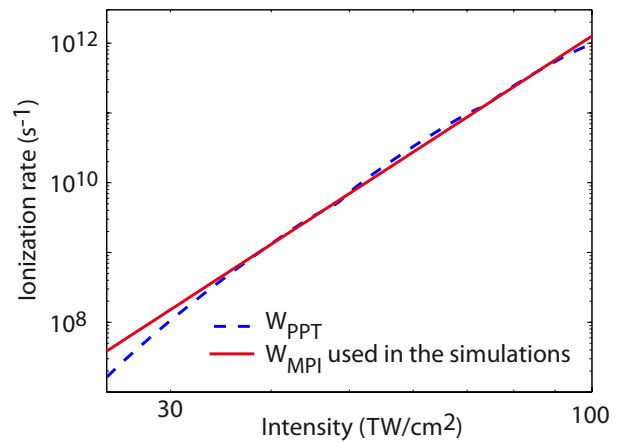}
  \end{center}
   \caption{(Color online) Ionization rate at 800 nm calculated with the PPT formulation (blue dashed line) as a function of intensity and the effective multiphoton rate following a Nth-order ($N=7.5$)  intensity power law used in the present work (red solid line). }
  \label{PPT_versus_MPI}
\end{figure}

The nonlinear refractive index expansion given in Eqs. \ref{delta_n_x} and \ref{delta_n_y} is consistent  with  the expression that has been used  in Refs. \cite{Loriot17_2009,Loriot18_2010} in order to define the  $n_{2j}$ values  from the birefringence measurements. 
This expression relies on the relation between the parallel and perpendicular component of the two first terms $n_2$ and $n_4$  derived  by Arabat and Etchepare   \cite{Arabat10_1993} and  generalized at any order by following the same approach \cite{Loriot17_2009}.  More recently, an analytical derivation including up to the $n_8$ term has shown that are our formula is not  strictly valid,  although the error introduced  on the  reported high-order Kerr terms is only a few percent \cite{Stegeman19_2011} (which remains within the experimental error of our measurements). For consistency, since  the present birefringence calculations are based on the Kerr index  values reported in Ref. \cite{Loriot18_2010},  we use the same expansion as previously.

For a pump and probe field time delayed by $\tau$ and  linearly polarized as depicted in Fig. \ref{setup}, the initial electric field {writes}:
\begin{eqnarray}
\mathcal{E}_{x}(z=0)&=&G(y)[\mathcal{E}_{\textrm{pump}}F_{\textrm{pump}}(t)H_{\textrm{pump}}(x)+\\
&+&\frac{1}{\sqrt{2}}\mathcal{E}_{\textrm{probe}}F_{\textrm{probe}}(t-\tau)H_{\textrm{probe}}(x)]\nonumber\\
\mathcal{E}_{y}(z=0)&=&\frac{1}{\sqrt{2}}\mathcal{E}_{\textrm{probe}}F_{\textrm{probe}}(t-\tau)G(y)H_{\textrm{probe}}(x)
\end{eqnarray}

with

\begin{eqnarray}
G(y)&=&e^{-\frac{y^2}{\sigma_y^2}}e^{-ik_0\left(R_z-\sqrt{R_z^2-y^2}\right)},\nonumber\\
H_{\textrm{pump}}(x)&=&e^{-\frac{(x+x_0)^2}{\sigma_x^2}}e^{i\Phi_{\textrm{pump}}(x)},\nonumber\\
H_{\textrm{probe}}(x)&=&e^{-\frac{(x-x_0)^2}{\sigma_x^2}}e^{i\Phi_{\textrm{probe}}(x)},\nonumber\\ F_{\textrm{pump}}(t)&=&e^{-\frac{t^2}{\sigma_t^2}},\nonumber\\
\Phi_{\textrm{pump}}(x)&=&-k_0\left(R_z-\sqrt{R_z^2-(x+x_0)^2}+\sin\theta(x+x_0)\right),\nonumber\\
\Phi_{\textrm{probe}}(x)&=&-k_0\left(R_z-\sqrt{R_z^2-(x-x_0)^2}-\sin\theta(x-x_0)\right),\nonumber\\
F_{\textrm{probe}}(t)&=&e^{-\frac{t^2}{\sigma_t^2}}\textrm{e}^{i\Delta\omega t},\nonumber\\
\sigma_x&=&\sigma_y=\sigma_0\sqrt{1+\frac{z_0^2}{z_r^2}},\nonumber\\
R_z&=&z_0\left(1+\frac{z_r^2}{z_0^2}\right), z_r=\frac{\pi\sigma_0^2}{\lambda_0}, x_0=-z_0\tan(\theta),\nonumber\\
\mathcal{E}_{\textrm{pump}} &=&\sqrt{\frac{2}{\pi}\frac{P_{\textrm{pump}}}{\sigma_x\sigma_y}}, P_{\textrm{pump}} =\sqrt{\frac{2}{\pi}}\frac{E_{\textrm{pump}}}{\sigma_t},\nonumber\\
\mathcal{E}_{\textrm{probe}} &=&\sqrt{\frac{2}{\pi}\frac{P_{\textrm{probe}}}{\sigma_x\sigma_y}}, P_{\textrm{probe}} =\sqrt{\frac{2}{\pi}}\frac{E_{\textrm{probe}}}{\sigma_t},\nonumber
\end{eqnarray}
where $R_z$ is  the curvature radius and $z_r$ is the Rayleigh length. The initial conditions were chosen to be as close as possible to the experimental ones. The pump beam waist  $\sigma_0$ is 40 $\mu$m (34 $\mu$m) in the HOK (PG) model case, $\theta$=3$^\circ$, $\lambda_0$=790 nm, $z_0$=-5 mm, and $\sigma_t$=85 fs. In the two last equations,  $E_{\textrm{pump}}$ ($E_{\textrm{probe}}$) is the \textrm{pump} (\textrm{probe}) energy and $P_{\textrm{pump}}$ ($P_{\textrm{pump}}$) is the pump (probe) peak power. Note that the \textrm{probe} energy was kept constant in all the simulations to $E_{\textrm{probe}}$=0.8 $\mu$J in order to match the experiment. The frequency detuning between the pump and probe pulse is $\Delta\omega$.

At the end of the propagation ($z$=10 mm), the probe beam does not overlap the pump beam anymore allowing the filtering of the former. The pure heterodyne signal obtained at a pump-probe delay $\tau$ is then evaluated for each spatial point by using a Jones matrix formalism describing the whole experimental detection setup and integrated over the entire spatiotemporal probe profile.

Note that, unlike the  model developed in Ref. \cite{Wahlstrand2011OPL}, the 3D+1 model intrinsically embeds the plasma grating effect without the need to add its contribution in the model. For comparison with the full model, which takes into account both plasma grating and HOK, the HOK model without plasma grating is obtained by setting the ionization cross-section $\sigma_N$ to 0, while the PG model is obtained by setting the HOK indices to 0.

Finally, one has to emphasize that, since the plasma grating model relies on the ionization process, the latter has to be modeled as good as possible. Such a modelling, which remains a challenging task, has been the subject of numerous theoretical works \cite{Keldysh65,PPT66,ADK86, Popruzhenko08}. In particular, it has been shown \cite{Talebpour99} that the ionization model first developed by Perelomov Popov and Terent'ev (PPT) reproduces well the experimental measurements performed in conditions close to the ones used in the present paper. However, the use of such a model leads to a dramatic increase of the computing-time and becomes prohibitive in 3D+1 calculations when systematic studies have to be performed like in the present work. Nevertheless, PPT ionization rates can be accurately fitted by an effective multiphoton law (with a nonlinearity N=7.5) in the intensity range 40-100 TW/cm$^2$ \cite{KasparianChin00}, as shown  in Fig. \ref{PPT_versus_MPI}. Such a nonlinearity, better suited for intensive 3D+1 calculations than the PPT formulation, has been confirmed by recent experimental measurements \cite{Wells2002,BejotPRA2013}.

\section{RESULTS AND DISCUSSION}
\label{results_section}
\subsection{One-color measurements}
\label{one-color_section}
\subsubsection{Fourier-transform limited (FTL) pulse}
\label{FTF}
 \begin{figure}[tb!]
  \begin{center}
    \includegraphics[keepaspectratio, width=9cm]{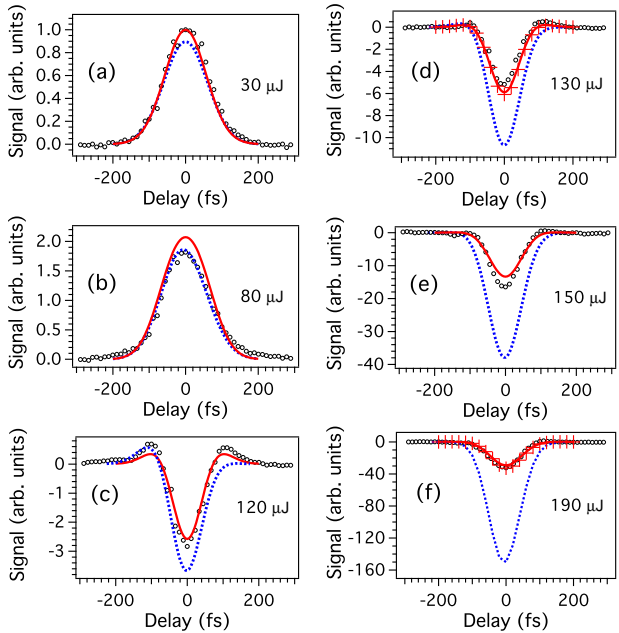}
  \end{center}
    \caption{(Color online) Pump energy dependence of the heterodyned birefringence signals (black circles) recorded in argon at a static pressure of (a-c) 400 and (d-f) 100 mbar versus the pump-probe delay. Data are compared to the 3D+1 numerical simulations based on PG (blue doted curves),  pure HOK model (red solid curves), and full HOK + PG model (red crosses). The Fourier-transform limited (FTL) pulse duration is 100 fs and the wavelength of the pump and probe is 790 nm. The beam waist in the PG model is  reduced by  17 \% compared to the measured value (see text). It is noted that all curves have been centered at 0 pump-probe delay for comparison. The peak pump intensities $I_{\textrm{pump}}$ (expressed in TW/cm$^2$) evaluated with numerical simulations are $I_{\textrm{pump}}$(HOK)$\simeq$0.37$\times E_{\textrm{pump}}$ and $I_{\textrm{pump}}$(PG)$\simeq$0.52$\times E_{\textrm{pump}}$, where $E_{\textrm{pump}}$ is the pump energy expressed in $\mu$J.}
  \label{int_dep}
\end{figure}
The one-color pump-probe measurements have been conducted with 790 nm pump and probe beam both delivered by a Ti:Sapphire amplifier without using the NOPA system presented in Fig. \ref{setup}. The birefringence signal recorded in argon with a heterodyne detection is shown with black circles in Fig. \ref{int_dep} for different input energies of the pump beam. All the experimental data are reported with the same scaling factor enabling thus a direct comparison between the different energies used. Each data point represents an averaging over 200 laser shots with a temporal sampling of 10 fs. In order to avoid significant nonlinear propagation effects, like intensity clamping, the gas pressure in the static cell was reduced to a low value reconcilable with an acceptable signal-to-noise ratio. As in the previous experiments \cite{Loriot17_2009,Loriot2011}, the compression of the laser pulses at the interaction region was checked by maximizing the signal. The result obtained at 30 $\mu$J [Fig. \ref{int_dep} (a)] is representative of the linear regime. It is featured by a linear variation of the signal near and below this energy and a temporal shape consistent with a birefringence $\Delta n$ associated with a pulse profile of the 100 fs-pump pulse. For laser energies close to and higher than 120 $\mu$J [Fig. \ref{int_dep} (c)]  the linear dependence of the signal on the pump energy is lifted and sign inversion is observed. The signal  shown in Fig. \ref{int_dep}(c) is typical of this inversion   and has been already reported in atomic or molecular gas species close to 25 TW/cm$^2$\cite{Loriot17_2009,Loriot18_2010,Loriot2011}. At higher energy values, the signal is dominated by a negative, rather symmetric, temporal profile reaching a minimum value close to the zero delay.
 \begin{figure}[tb!]
  \begin{center}
    \includegraphics[keepaspectratio, width=9cm]{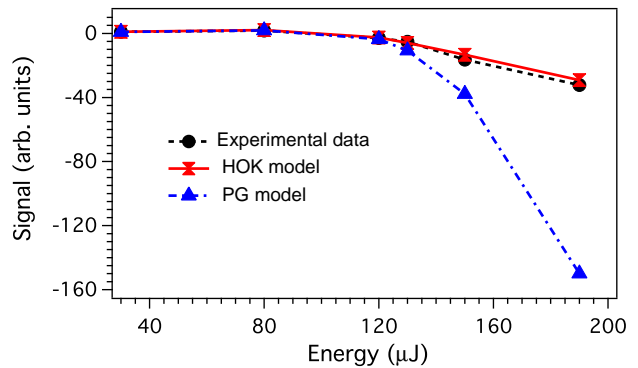}
  \end{center}
   \caption{(Color online) Energy dependence of the  birefringence signal (see Fig. \ref{int_dep}) measured  near the zero pump-probe delay compared to both models.}
  \label{peaksignal}
\end{figure}

A comparison between the 3D+1 numerical simulations and the experimental data is also presented in Fig. \ref{int_dep}. The results based on the HOK model alone, i.e.,  without accounting for the ionization process, are shown with red solid curves. As shown the data are fairly well reproduced by the HOK model using the Kerr indices measured in Refs. \cite{Loriot17_2009,Loriot18_2010}.
Since the model does not account for ionization, the negative part of the signal results from the negative high-order Kerr terms $n_4$ and $n_8$ [see Eqs. \ref{delta_n_x} and \ref{delta_n_y}] that are the dominant contributions to the nonlinear refractive index above 120 $\mu$J.  The influence of the ionization, and hence the  contribution of the  plasma grating to the birefringence measurements, has been assessed by calculating the theoretical signals without HOK. %, using the same input beam size as previously.
Because the ionization rate of Ar predicts a balance of the positive Kerr ($n_2$) by the PG effect beyond the energy employed in the experiment, the simulations have been performed with a beam waist decreased by 17 \% with respect to the measured value so as to allow a  comparison with the experimental data. It should be emphasized that this adjustment does not impact the energy dependence of the signal. The results are shown with blue doted curves.  Comparison between experimental data and PG model clearly indicates that the latter fails to reproduce the nonlinear dependence of the observed effect above the inversion. This is shown also in Fig. \ref{peaksignal} where the dependence of the recorded signal on the laser energy is depicted along with the corresponding numerical simulations results for both models. Under the present experimental conditions, the ionization rate of Argon  calculated in Fig. \ref{PPT_versus_MPI} approximately scales like the seventh power of the intensity  between 40 and   100 TW/cm$^2$ \cite{KasparianChin00,BejotPRA2013} and  is  negligible  for lower intensities. The magnitude of the PG effect being proportional to the ionization rate requires that the resulting birefringence obeys the same intensity dependence \cite{Wahlstrand2011OPL,Wahlstrand85_2012}. On the contrary, the birefringence signal above 120 $\mu$J changes like the fourth power of the applied energy, which is properly captured by the HOK model with $n_8$ being the dominant term at  large intensity. Simulations including both HOK and PG effect have also been implemented using the experimental parameters of Fig. \ref{int_dep}. No significant difference with the pure HOK model was observed as shown for instance in  Figs. \ref{int_dep} (d) and (f). Finally, {Fig. \ref{clamping} shows that the intensity clamping resulting from nonlinear propagation effect is insignificant (less than 5$\%$), even in the PG model where the intensity is larger than in the HOK simulations.  The previous results confirm that under the present conditions the PG acts as a marginal term in the present birefringence experiment.
 \begin{figure}[tb!]
  \begin{center}
    \includegraphics[keepaspectratio, width=9cm]{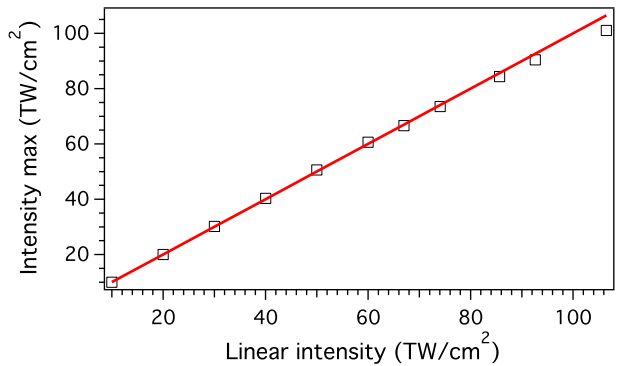}
  \end{center}
   \caption{(Color online) Peak intensity (black squares) calculated with the help of 3D+1 simulations in the case of the PG model as a function of the  intensity expected in the linear propagation regime. The linear regime (red solid curve) is shown to help the eye.}
  \label{clamping}
\end{figure}

\subsubsection{Chirped pulse}
\label{chirpedpulse}
The birefringence induced by the plasma grating results from the accumulation of electrons produced all along the pulse duration. Due to its delayed nature, the negative phase shift of the PG combined with the positive $n_2$ Kerr index produces an asymmetric signal around the intensity of inversion where both effect counteract each other \cite{Wahlstrand2011OPL,Wahlstrand213}. No significant asymmetric profile was observed neither in the previous experiment \cite{Loriot17_2009,Loriot2011} nor in the present one [see for instance Fig. \ref{int_dep} (c)]. However, it was recently  suggested that the absence of asymmetry could be the result of the the PG effect when considering temporally chirped pulses \cite{Wahlstrand213}. Although all our experiments so far were performed for compressed pump and probe beams, a set of data recorded with chirped pump and probe  pulses around the region of inversion are compared to numerical simulations in Fig. \ref{chirp}. The linear chirp parameter has been characterized  by  SPIDER [Figs. \ref{chirp} (a-c)] and autocorrelation [Figs. \ref{chirp} (d, e)] measurements. It is noticed  that $\pm$2150 fs$^2$ corresponds to a measured pulse duration of 160 fs. In order to compensate for the intensity drop introduced by the pulse stretching compared with the FTL pulse, the energy of the chirped pulse has been increased so as to recover the signal near the inversion region. The structural shape of the signals slightly differ with each other due to the high sensitivity of the effect with respect to the intensity. % Because  the structural shape of the signal is very sensitive to the input energy near the inversion, it was not possible to recover the exact same shape as the FTL pulse [see e.g. Fig. \ref{chirp}(f)].
Nevertheless, one can see that the symmetry of the data is well conserved even for the largest chirp employed. On the contrary, the simulations presented in the same figure show that the  structural shape  of the PG signal is sensitive to the chirp \cite{Wahlstrand213}. Although, near the inversion the  FTL-pulse signal  is featured by an  asymmetric shape (see Fig. \ref{chirp} (b)], the latter can be reduced  when using  large chirps as shown for instance in Figs. \ref{chirp} (d, e). The fact  that the symmetry of the observed signals  is preserved while the pulses are chirped and also  that the HOK model does not predict any symmetry reversal of the signal allows to exclude the chirp as a possible artifact from our measurements. Finally, it should be pointed that large chirps, as those  responsible for symmetric PG signal \cite{Wahlstrand213}, lead  at  constant energy level to a dramatic drop of intensity  and consequently to a significant decrease of the birefringence signal above the inversion threshold. This along  with a temporal broadening  of the signal could have  been straightly noticed  in  the previous studies \cite{Loriot17_2009,Loriot2011}  so that any unforeseen  significant chirp would have  been   corrected. 
\begin{figure}[tb!]
  \begin{center}
    \includegraphics[keepaspectratio, width=9cm]{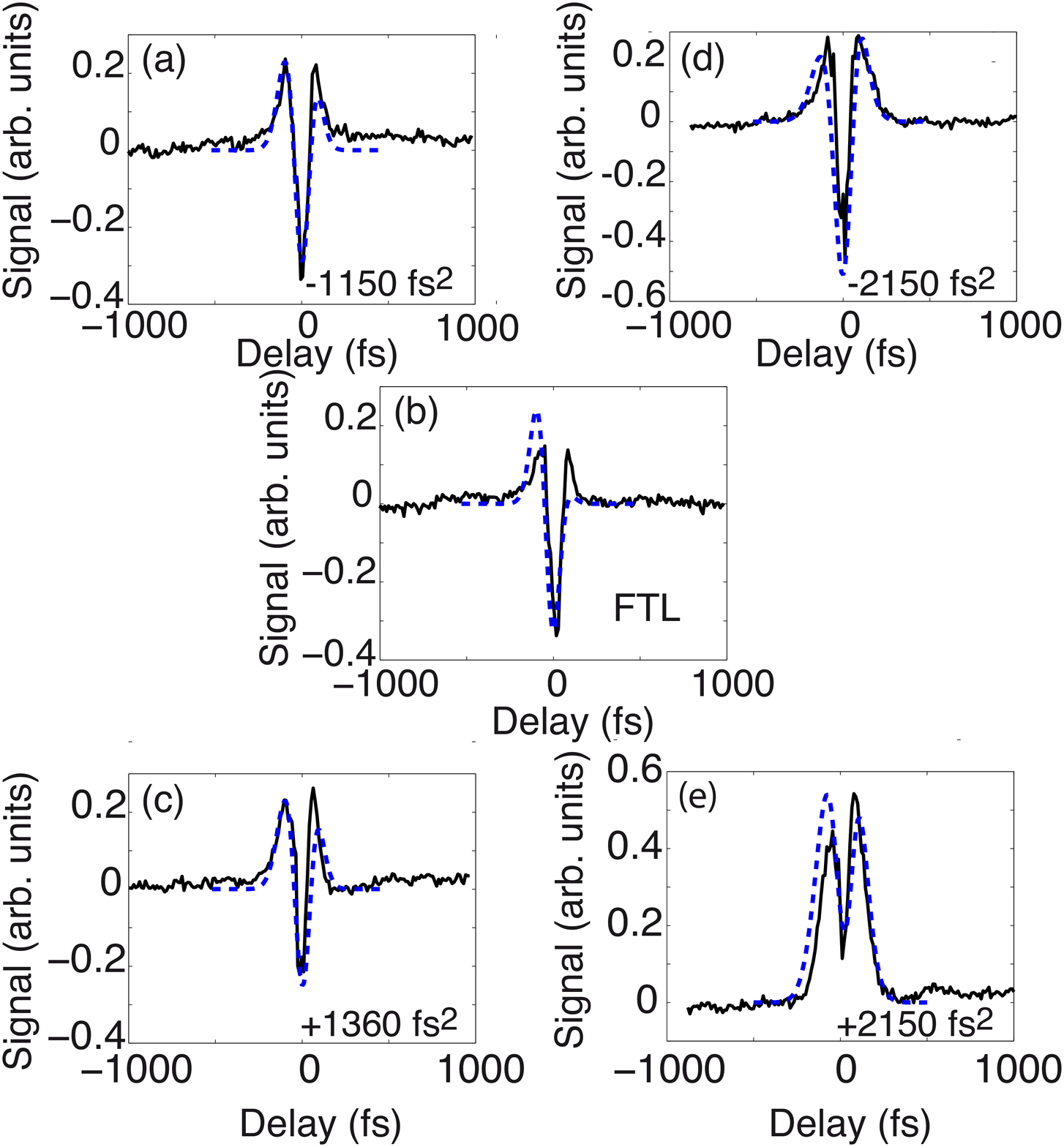}
  \end{center}
    \caption{(Color online) Comparison between heterodyned birefringence signals (black solid curves) recorded with  (b)  Fourier-transform limited (FTL)  pulses  and chirped pulses  with different linear chirps: (a) -1155 fs$^2$, (c) +1360 fs$^2$, (d) -2150 fs$^2$, and (e) +2150 fs$^2$. Numerical simulations (blue dashed curves) based on the  plasma-grating effect. }
  \label{chirp}
\end{figure}

\subsection{Two-color measurements}
\label{two-color_section}
\begin{figure}[tb!]
  \begin{center}
    \includegraphics[keepaspectratio, width=8cm]{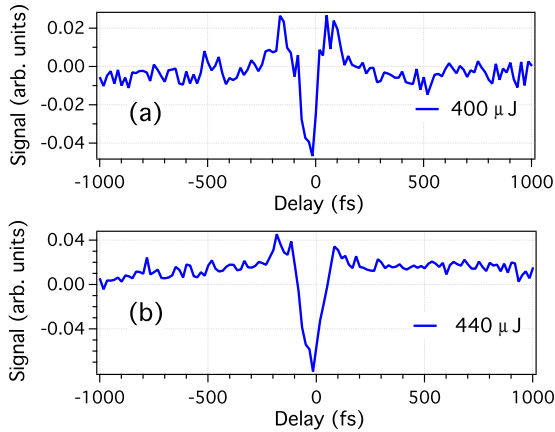}
  \end{center}
    \caption{(Color online) Two-color pump-probe heterodyne birefringent signals recorded in 100 mbar of argon with a pump input energy of (a) 400 and (b) 440 $\mu$J.}
  \label{two_color}
\end{figure}Recently, a spectral analysis of the nonlinear polarization calculated with time-dependent Schr\"odinger equation \cite{Bejot110_2013} revealed that the saturation and inversion of the nonlinear refractive index mainly occurs within the bandwidth of the pump pulse in agreement with a recent observation \cite{Odhner2012}. This is consistent with single-color pump-probe experiment, which exhibits refractive index saturation and inversion \cite{Loriot17_2009,Loriot2011,Odhner2012}, and two-color experiments using non overlapping spectra that do not \cite{Wahlstrand2011PRL,Wahlstrand2012PRL,Odhner2012}. However, the results presented in this section show that the sign inversion can also be produced with two-color non-overlapping spectra provided that much larger intensity is used compared to the one-color scheme.

The two-color studies have been performed with a 100 fs probe beam delivered by a non-collinear optical parametric amplifier (NOPA) inserted in the probe beam line, as shown in Fig. \ref{setup}. The data of Fig. \ref{two_color} have been recorded for a pump and probe wavelength set at 790 and 740 nm, respectively. The birefringence signals presented in this figure are similar to those of Figs. \ref{int_dep} (c-d) or Fig. \ref{chirp} (b), except that the  energy had  to be increased by a factor of $\sim$3 in order to observe the inverted signal. So far this is the first observation of negative birefringence observed with non degenerate frequencies. Experiments performed with a larger detuning of the probe frequency, e.g., with 650 nm probe wavelength, did not reveal any inversion of the signal up to 800 $\mu$J.

\begin{figure}[tb!]
  \begin{center}
    \includegraphics[keepaspectratio, width=9cm]{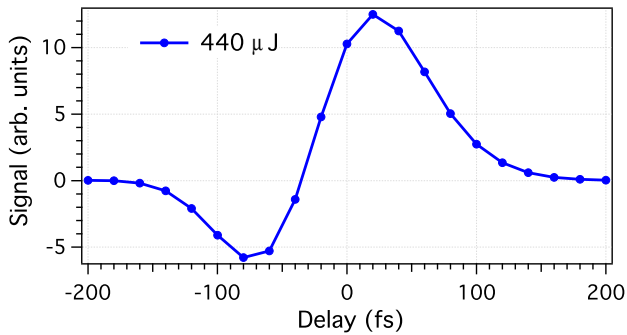}
  \end{center}
    \caption{(Color online) Two-color simulation of the plasma grating-induced birefringence for the input energy and pressure corresponding to Fig. \ref{two_color}(b).}
  \label{two_color_simul}
\end{figure}
\begin{figure}[tb!]
  \begin{center}
    \includegraphics[keepaspectratio, width=9cm]{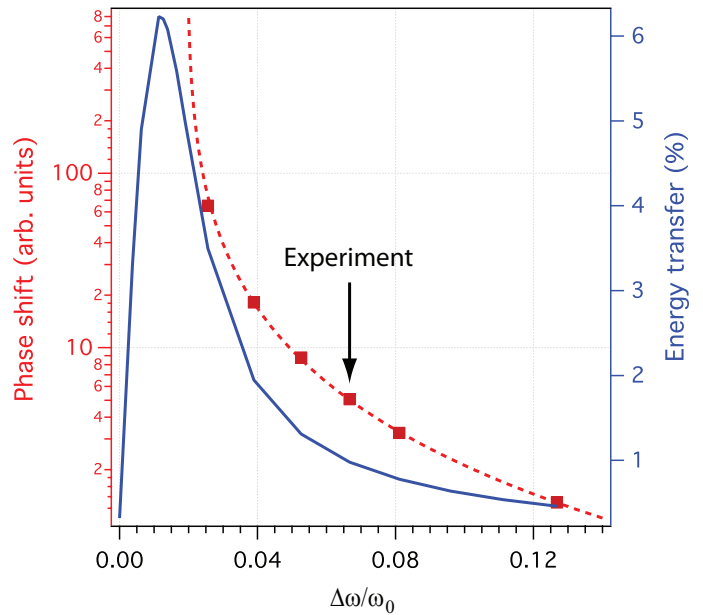}
  \end{center}
    \caption{(Color online) Plasma grating phase shift  calculated numerically (left scale, red squares), its associated fit $\propto1/\left(\alpha+\beta \Delta \omega^2\right)$ (red dashed curve), and energy transfer (right scale, blue solid curve) calculated as a function of the frequency difference $\Delta \omega$ between the pump and probe beam. The arrow indicates the frequency difference of the experiment. }
  \label{PG_versus_frequency}
\end{figure}

To point out the influence of the plasma grating in the two-color experiment, a 3D+1 numerical simulation corresponding to the experimental conditions of Fig. \ref{two_color} was performed without HOK terms, but, as in the one-color case, with a beam waist reduction of 17\%. The result shown in Fig. \ref{two_color_simul} reveals that the PG effect also occurs for non degenerate frequencies, but it is much  reduced, therefore larger energies are required for its observation. This is an unexpected result considering that as relying on optical interference, the PG effect is unlikely to take place with two-color non-overlapping spectra \cite{Odhner2012,Wahlstrand213}. In order to explain this apparent contradiction, let us recall that in two-color optical grating the difference between the two beam frequencies leads to a transverse shift of the optical fringes during the pulse. When the frequency detuning exceeds the pulse bandwidth, the spatial shift  covers a distance larger than  the fringe spacing of the optical grating  during the pulse duration. Since ionization occurs within the whole pulse duration, the accumulated charges combined with the spatial shift of the optical grating should lead to the obliteration of the fringe contrast at the end of the pulse. However, rare gas ionization at 790 nm is a highly nonlinear process which mainly occurs around the maximum of the pulse, over a temporal window much shorter than its temporal width. During this short time interval, the contrast can be maintained even for frequency differences exceeding the pulse bandwidths. The frequency difference enabling a contrasted grating is therefore inversely proportional to the ionization window. Figure \ref{PG_versus_frequency} shows the birefringence resulting from the PG effect calculated for different pump-probe frequency differences $\Delta \omega$. The effect is well fitted by a $1/\left(\alpha+\beta \Delta \omega^2\right)$ law as shown in Fig. \ref{PG_versus_frequency}.

As in the one-color experiment, we can see that the birefringence signal of the two-color experiment does not match the calculated PG effect. In particular, a salient difference is observed between  the temporal profiles. The pure PG effect in the non-degenerate frequencies case generates negative dispersion-like temporal shape centered on the zero delay, with a minimum (maximum) located at negative (positive) delay. When combined with the positive $n_2$ Kerr contribution, it results in a birefringence effect with  a temporal profile such as in Fig. \ref{two_color_simul}, which is totally inconsistent with the present  observation. It was checked numerically that this disagreement could not result from %the shape difference between the PG model prediction and the experimental result cannot be explained by
 a  residual chirp of the pump and/or the probe beam. Finally, according to our numerical results, the high energies used in the measurements of Fig. \ref{two_color}  lead to the clamping of the pump intensity by a factor of $\sim$2. Nevertheless, since our numerical simulations take into account the clamping phenomenon, the difference between PG model predictions and the experimental results cannot be assigned  to the later.

\begin{figure}[tb!]
  \begin{center}
    \includegraphics[keepaspectratio, width=9cm]{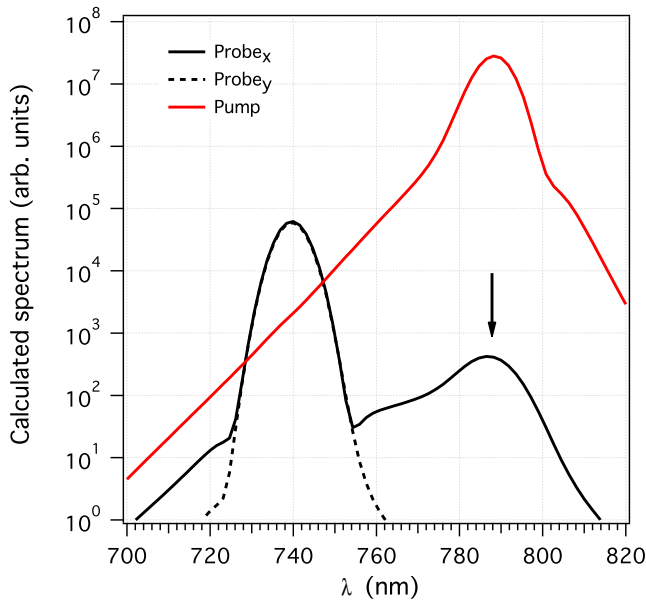}
  \end{center}
    \caption{(Color online) Polarization resolved pump and probe spectrum calculated after propagation though the Ar medium. The subscripts $x$ and $y$ refer to the polarization direction of the fields. Same conditions as in Fig. \ref{two_color_simul}.  }
  \label{energy_transfer_theo}
\end{figure}
A plasma grating can also be responsible for energy transfer between the pump and probe beam in  two-color experiments \cite{Yang_grating2013}, or in  one-color chirped-pulse experiments \cite{Mysy_grating2013}. It is known that the phase shift produced by the two-beam coupling is maximized for degenerate frequencies \cite{Boyd2008}. On the contrary, the energy transfer does not reach its optimal value for degenerate frequencies but for a particular frequency detuning that depends on the characteristic time needed for  building up the grating through ionization. Because of the spatial and temporal modulation of the grating, energy flow occurs from the pump to the probe beam, and \textit{vice versa} depending on the sign of $\Delta \omega$. Both result in an energy exchange around the pump and probe frequency. For $\omega_{\textrm{probe}}>\omega_{\textrm{pump}}$, the simulations of Fig. \ref{energy_transfer_theo} show that the energy flow is directed toward the probe beam, as evidenced by the energy bump centered around 790 nm and indicated by an arrow on the probe spectrum. This $\sim$1\% of energy transfer is confirmed by the observation. Figure \ref{energy_transfer_exp} shows the probe beam spectrum recorded with and without gas. Because the probe- to pump-energy ratio is less that $10^{-3}$, the relatively low energy loss experienced by the pump is not noticeable on the pump spectrum.
 \begin{figure}[tb!]
  \begin{center}
    \includegraphics[keepaspectratio, width=8cm]{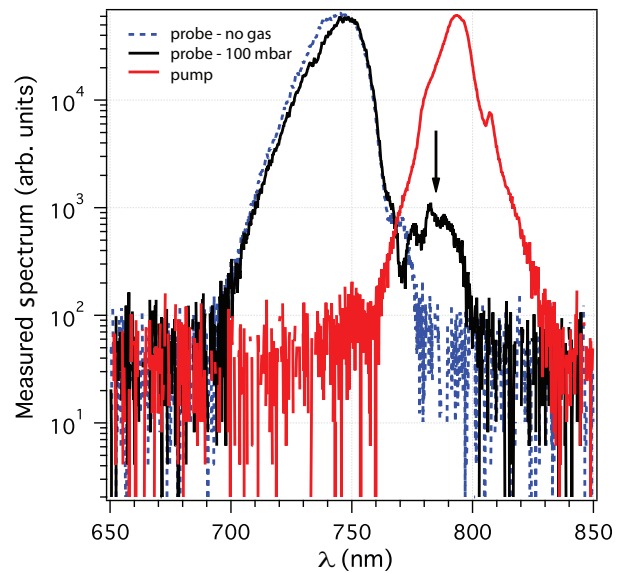}
  \end{center}
   \caption{(Color online) Normalized spectra measured at the medium exit. The pump energy is 440 $\mu$J.}
  \label{energy_transfer_exp}
\end{figure}

It is emphasized that our pump-probe measurements are not affected by neither polarization dependent energy transfer nor dichro\"{\i}sm experienced by the probe as long as the ratio between the two polarization components remains close to unity. More particularly, using the Jones matrix formalism and assuming that the probe electric field at the end of the interaction has the general form
 \begin{eqnarray}
 E_x&=&\sqrt{T}\textrm{e}^{i\Phi}\nonumber\\
 E_y&=&1,
 \end{eqnarray}
where $\Phi$ is the nonlinear phase difference between the two polarization axes and $T$ is the intensity transmission along the $x$ dimension ($T$ can be either lower than unity in the case of losses or higher than unity in the case of gain), one can show that  the full heterodyne signal  measured with  the present technique is given by

\begin{equation}
S_{\textrm{heterodyne}}\propto 2\sqrt{T}\sin\Psi_{\textrm{L.O.}}\sin{\Phi},
\end{equation}
where $\Psi_{\textrm{L.O.}}$ is the phase of the local oscillator introduced in Eq. \ref{local osci}.
 
According to our 3D numerical model the energy transfer, as predicted by the PG model, is reproduced well from our experimental results. However as mentioned before, the latter effect unambiguously cannot solely describe the imparted phase to the probe beam, i.e. the birefringence signal, as shown from the comparison of figures 7 and 8. 

\section{CONCLUSION}
\label{conclusion_section}
Concluding, the active debate concerning the impact of the plasma grating (PG) effect on the nonlinear dynamics of optically induced birefringence is revisited by a systematic and thorough study in argon gas. New experimental data are presented at two different configurations concerning the laser wavelength, i.e., the degenerate and the non-degenerate case and the results are compared to both HOK and PG models with the help of full 3D+1 numerical simulations.

In the degenerate configuration, the dependence of the induced birefringence on the power and the chirp of the laser pulse was examined. The power dependence measurements confirmed the sign inversion of the signal above a specific energy threshold and showed that for laser intensities well above the inversion it is proportional to the fourth power of the applied intensity, in full agreement with the HOK model. On the contrary, the PG effect is not adequate to account totally for the reported observations under the applied experimental conditions. In respect with the dependence on the chirp the symmetric peak profiles of the recorded signals for both chirped and FTL pulses further corroborate to this conclusion.

In the two non-degenerate beams configuration, it was shown, for the first time, that a saturation and a sign inversion of the nonlinear refractive index also take place. This sign inversion occurs at about 3 times higher intensity than in the degenerate case suggesting that the high-order indices are frequency-dependent. Moreover, the predicted birefringence profile from the  PG model is found  inconsistent with the recorded one while the predicted energy transfer (1\%) is in close resemblance with the experimental findings, making thus the impact of PG marginal also in this case. At this point it is emphasized that, besides its failure to reproduce our results, PG effect is still valid in the non-degenerate regime. The proposed interpretation for its occurrence is based on the fact that atomic ionization is a very fast process and takes place within a short time window around the peak intensity of the laser pulse. Thus, even in a non-degenerate beams scheme, it can be shown that despite the transverse shift of the optical fringes during the pulse duration, a contrasted grating remains inducing in turn a phase shift. Finally, albeit all the above experimental and theoretical data  confirm the validity of using the HOK as an appropriate approach to  the atomic response at moderate laser intensities, we do not overlook the fact that further studies, in both theory and experiment, are needed so as to provide the full physical mechanism behind it.

\bibliography{Biblio_PRA_HOKE}
\acknowledgments
This work was supported by the Conseil R\'egional de Bourgogne (PARI and FABER programs), the CNRS, and the Labex ACTION program (contract ANR-11-LABX-01-01). P.B. thanks the CRI-CCUB for CPU loan on its multiprocessor server.\vfill

\end{document}